\documentclass[aps,prl,epsf,twocolumn,final]{revtex4}

\usepackage{graphicx}
\usepackage{latexsym}
\usepackage{amssymb}
\usepackage{amsxtra}
\usepackage[usenames,dvipsnames]{color}
\usepackage{mathptmx}
\usepackage{soul,color}
\sethlcolor{green} \setulcolor{blue} \setstcolor{red}

\begin{document}
\title{Laser Wakefield Acceleration Using Wire Produced Double Density Ramps}
\author{M. Burza,$^1$ A. Gonoskov,$^{2,3}$ K. Svensson,$^1$  F. Wojda,$^1$ A. Persson,$^1$ M. Hansson,$^1$\\ G. Genoud,$^1$ M. Marklund,$^2$ C.-G. Wahlstr\"om$^1$ and O. Lundh$^1$}
\affiliation {\vskip 0.30cm $^1$ Department of Physics, Lund University, P.O. Box 118, SE-221 00 Lund, Sweden\\
$^2$ Department of Physics, Ume{\aa} University, SE-901 87 Ume\aa, Sweden\\
$^3$ Institute of Applied Physics, Russian Academy of Sciences, 46 Ulyanov Street, Nizhny Novgorod 603950, Russia}
\date{\today}

\begin{abstract}
A novel approach to implement and control electron injection into the accelerating phase of a laser wakefield accelerator (LWFA) is presented. It utilizes a wire, which is introduced into the flow of a supersonic gas jet creating shock waves and three regions of differing plasma electron density. If tailored appropriately, the laser plasma interaction takes place in three stages: Laser self-compression, electron injection and acceleration in the second plasma wave period. Compared to self-injection by wavebreaking of a nonlinear plasma wave in a constant density plasma, this scheme increases beam charge by up to one order of magnitude in the quasi-monoenergetic regime. Electron acceleration in the second plasma wave period reduces electron beam divergence by $\approx 25\,$\%, and the localized injection at the density downramps results in spectra with less than a few percent relative spread.
\end{abstract}

\maketitle

Plasma-based laser-driven electron accelerators can produce strong longitudinal fields, $\sim 100\,$GV/m, in the collective electron oscillations in the wake of an intense laser pulse \cite{Tajima1979}. This gives an advantage over conventional accelerators using RF cavities regarding the relatively compact high-power table-top laser systems readily available \cite{Backus1998}.

In most experiments, injection of electrons into the accelerating structure relies on breaking of the plasma wave, which can thus self-inject electrons. This scheme is rather simple and quasi-monoenergetic beams have been produced \cite{Geddes2004, Faure2004, Mangles2004}. Electron beams of low spectral spread and divergence are necessary for these accelerators to be attractive for applications \cite{Nakajima2008, Schlenvoigt2008, Fuchs2009}. However, the wavebreaking process is highly nonlinear, and in order to achieve higher quality beams, means to control the injection process are required. Both, the amount of charge and the time of electron injection from the background plasma into the accelerating and focusing phase of the wakefield are crucial \cite{Tzoufras2008, Rechatin2009, Rechatin2009a}. Here, self-injection \cite{Kostyukov2009, bulanov.prl.1997, zhidkov.pop.2004} is inferior to most schemes with external injection control, such as colliding pulse techniques \cite{Esarey1997, Malka2009, beck.njp.2011, Faure2006}, ionization injection \cite{Pak2010, McGuffey2010} or gradients in plasma electron density \cite{Bulanov1998, suk.prl.2001, Geddes2008, Gonsalves2011}, which are used in this experiment. At the downwards gradient the plasma wavelength increases rapidly, the plasma wave breaks and electrons are trapped.

Shock waves resulting in very abrupt density transitions, have been produced previously with a knife edge introduced into a supersonic gas flow  \cite{Schmid2010}. By this, a well defined shock wave and a density downwards gradient is provided on the laser axis. Alternatively, an auxiliary pulse produced an electron depleted region by formation of an ionization channel followed by hydrodynamic expansion \cite{Faure2010, Hsieh2006}. Our experiment relates to these, as plasma densities are modulated on the laser axis to control injection externally.

We present a novel, staged, three step, laser wakefield accelerator that utilizes a thin wire crossing the supersonic flow of a gas jet. In this scheme, extremely sharp density transitions and shock waves facilitate gradient injection. These transitions may be of only some microns length \cite{Kermode2006}. In the first stage, comprised of a constant plasma density prior to reaching the first shock front, the pulse propagates and may thus match itself to the plasma conditions by relativistic self-focusing, self-modulation and temporal compression, with only a negligible amount of charge being  trapped. After a variable length, adjustable by the wire position along the optical axis, the laser pulse reaches the second stage, where the first shock front, originating from the wire, in combination with the subsequent expansion fan produces gradients that enable injection. During the transition to the third stage, the plasma density increases from the density-diluted region right above the wire to the final constant density region. The plasma wavelength shrinks rapidly, which under certain circumstances enables a controlled charge transfer of the previously injected electrons from the first into the second plasma wave period. This mechanism, which is driven by inertia, may in addition have a filtering effect on the previously injected charges. The dominant process however, is a new injection at the second shock front. The proceeding constant density plasma is finally utilized for electron acceleration driven by the already matched laser pulse.

\begin{figure}
\includegraphics[width=\columnwidth]{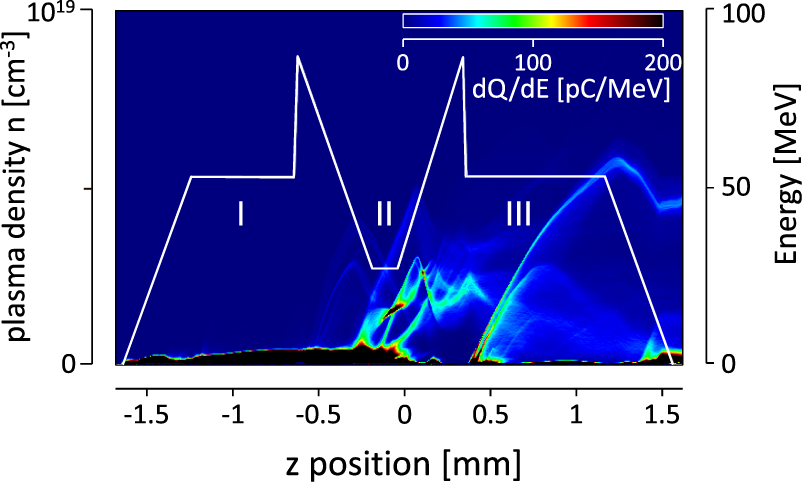}
\caption{\textcolor{black}{Electron energy distribution as a function of the laser pulse $z$-position and plasma density distribution (white curve).}} \label{sima}
\end{figure}

The experiment was conducted at the Lund Laser Centre, Sweden, where a Ti:Sa CPA laser system provided pulses at $800\,$nm central wavelength with $42\,$fs duration and $1\,$J energy. The laser field is horizontally polarized. A deformable mirror and an $f=75\,$cm off-axis parabolic mirror facilitated a nearly diffraction limited focal spot, $0.7\,$mm above the orifice of a $3\,$mm diameter supersonic gas nozzle. The laser was focused at the boundary of the gas jet producing a spot with $15\,\mu$m diameter (intensity FWHM). A motorized holder positioned a wire above the nozzle but below the laser optical axis ($z$-axis). This produced three distinct plasma density regions for the laser interaction as schematized by the white broken line function in Fig. \ref{sima}. To tailor and model plasma electron densities for simulations, interferometric measurements were carried out using hydrogen at $9\,$bar backing pressure. It was found that the laser pulse initially encounters a region of approximately constant electron density (region I), which was determined to $6 \times 10^{18}\,$cm$^{-3}$. After $\approx 1\,$mm it encounters the first shock wave and a downwards gradient, followed by region II, where the plasma density is reduced to $3 \times 10^{18}\,$cm$^{-3}$ over $\approx 300\,\mu$m. After a second shock wave transition, region III is reached. Here the density is approximately the same as in region I and it is here the main acceleration takes place. Plasma densities scale linearly with backing pressure. Adjustments of wire position, thickness, Mach number and backing pressure tailor gradients, lengths and density ratios between region I and II to match the requirements for electron injection and laser guiding. Shock wave divergence angle and density ramps were found to be symmetric as long as the wire is $< 0.5\,$mm off the nozzle centre. Without wire, the plasma density is almost constant and comparable to that at the plateau regions I and III.

\begin{figure}[h]
	\centering
		\includegraphics[width=\columnwidth]{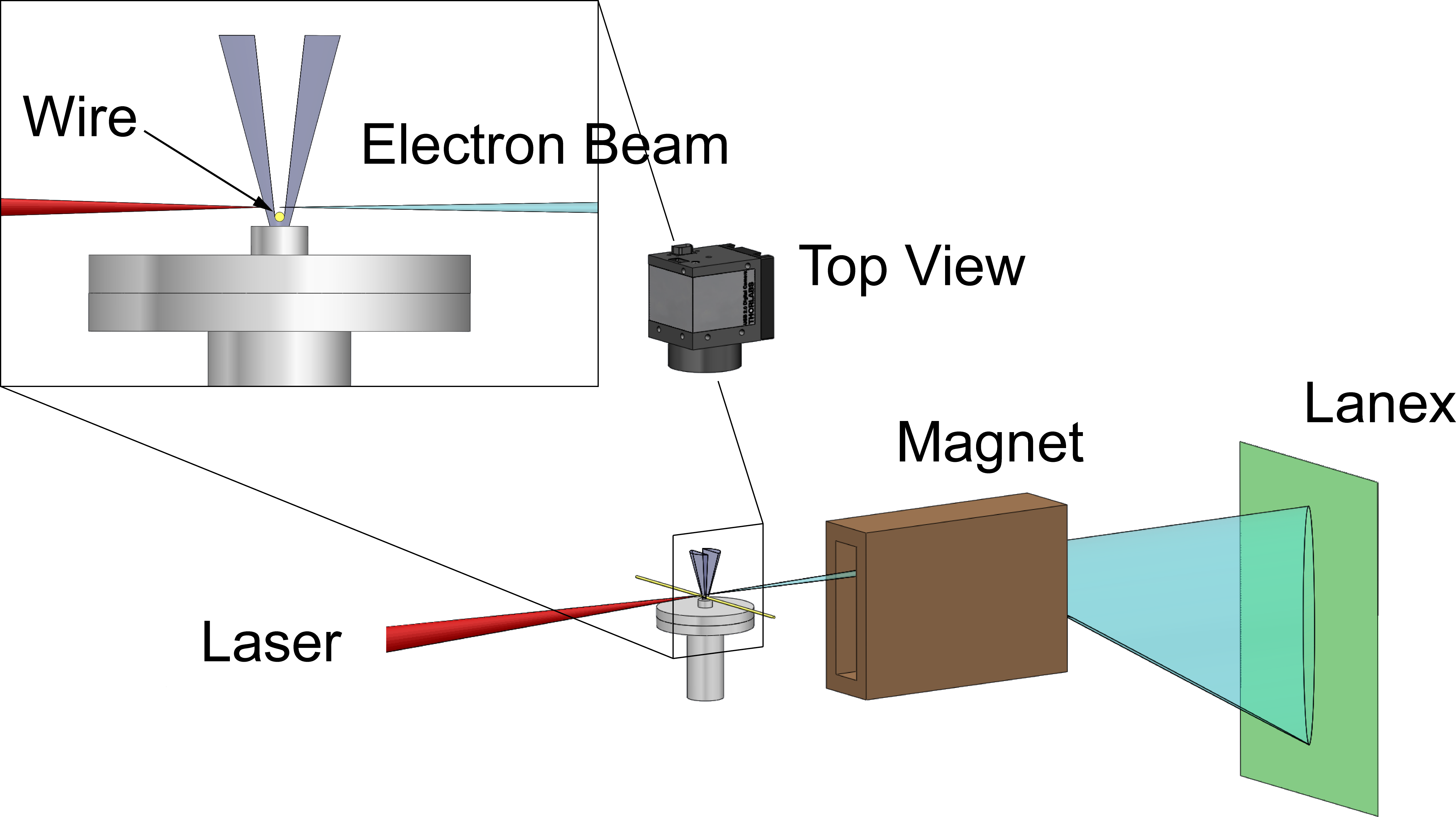}
	\caption{Experimental setup: The laser pulse enters from the left and impinges on the gas jet $0.7\,$mm above the nozzle. The wire is positioned $\approx 0.2\,$mm above the orifice. Top-view and a permanent magnet Lanex electron spectrometer serve as primary diagnostics. The position $z=0$ along the laser axis is centred above the nozzle.}
	\label{fig:interaction}
\end{figure}

As diagnostics served a top view camera and a permanent magnet electron spectrometer equipped with a Lanex screen (Kodak Lanex Regular), whose emission was recorded by a 16 bit CCD camera. Based on previous work \cite{Glinec2006, Buck2010} the electron spectrometer is calibrated in absolute charge. The setup is depicted in Fig. \ref{fig:interaction}.

Stainless steel wires with $300\,\mu$m, $200\,\mu$m, $50\,\mu$m and $25\,\mu$m diameter were tested, but only the latter two were found to trigger injection, with the clearly best performance with the $25\,\mu$m wire. Thicker wires inevitably increase the length and depth of the density-diluted region II, promoting diffraction and making it difficult to maintain a sufficiently focused laser pulse for region III.

Hydrogen and helium were both tested as target gas together with the wire but while hydrogen could deliver electron beams in more than $90\,$\% of the shots, helium was much less reliable with an optimized injection probability of less than $20$\%. This is in line with parallel studies investigating the influence of the target gas on beam quality and reliability in a constant density gas jet \cite{Svenssonitshapeetal.2012}. Thus, in the following, results obtained with hydrogen are presented.

A wire height scan revealed, that the probability for the production of electron beams increases with reduced distance to the optical axis. However, when closer than $0.65\,$mm the lifetime of the wire is reduced. As no improved performance on the production of electron beams could be observed in the range between $0.35\,$mm and $0.50\,$mm, the latter position was chosen.

A z-scan conducted with the $25\,\mu$m wire, $0.50\,$mm below the laser optical axis, and with a backing pressure below but close to the threshold for self-injection, revealed the sensitivity of the wire position along the optical axis on the production of electron beams (threshold is defined here as the constant plasma density resulting in beams with $<10\,$\% of the maximum charge observed during a pressure scan in the quasi-monoenergetic regime). This window was found to be $\approx 200\,\mu$m wide only. Outside this, the beam charge is comparable to the self-injection case without density modulation.

\begin{figure}[h]
	\centering
		\includegraphics[width=\columnwidth]{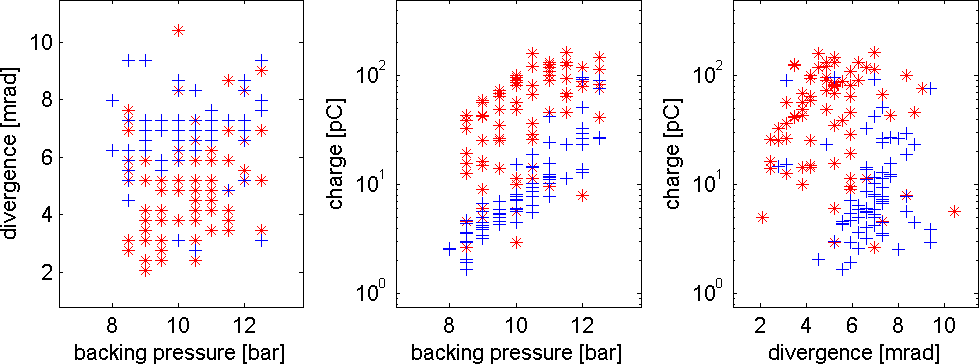}
	\caption{Comparison of divergence and charge of electron beams using hydrogen as target gas. Red stars represent shots with the wire $0.5\,$mm below the laser axis, and blue crosses represent LWFA with nonlinear wavebreaking and self-injection. Every data point corresponds to one shot. Failure rate with wire is below $5\,$\% and thus comparable to the wireless self-injection. Note the increased brightness indicated by the results in the figure to the right.}
	\label{fig:beamfeatures}
\end{figure}

Pressure scans were carried out at what was found to be the optimum spatial parameters, employing the $25\,\mu$m diameter wire at $0.50\,$mm distance to the optical axis and at a longitudinal $z$-position $0.07\,$mm from the nozzle centre towards the off-axis parabolic mirror. The wire injection scheme was found to be rather robust with regard to backing pressure. Below the self-modulated LWFA regime at $11\,$bar, the beam charge is increased by one order of magnitude, as illustrated in Fig. \ref{fig:beamfeatures}. With the wire, electron beam divergence is not affected by the overall plasma density but is on average only $75\,$\% compared to the self-injection case.\\

\begin{figure}
	\centering
		\includegraphics[width=\columnwidth]{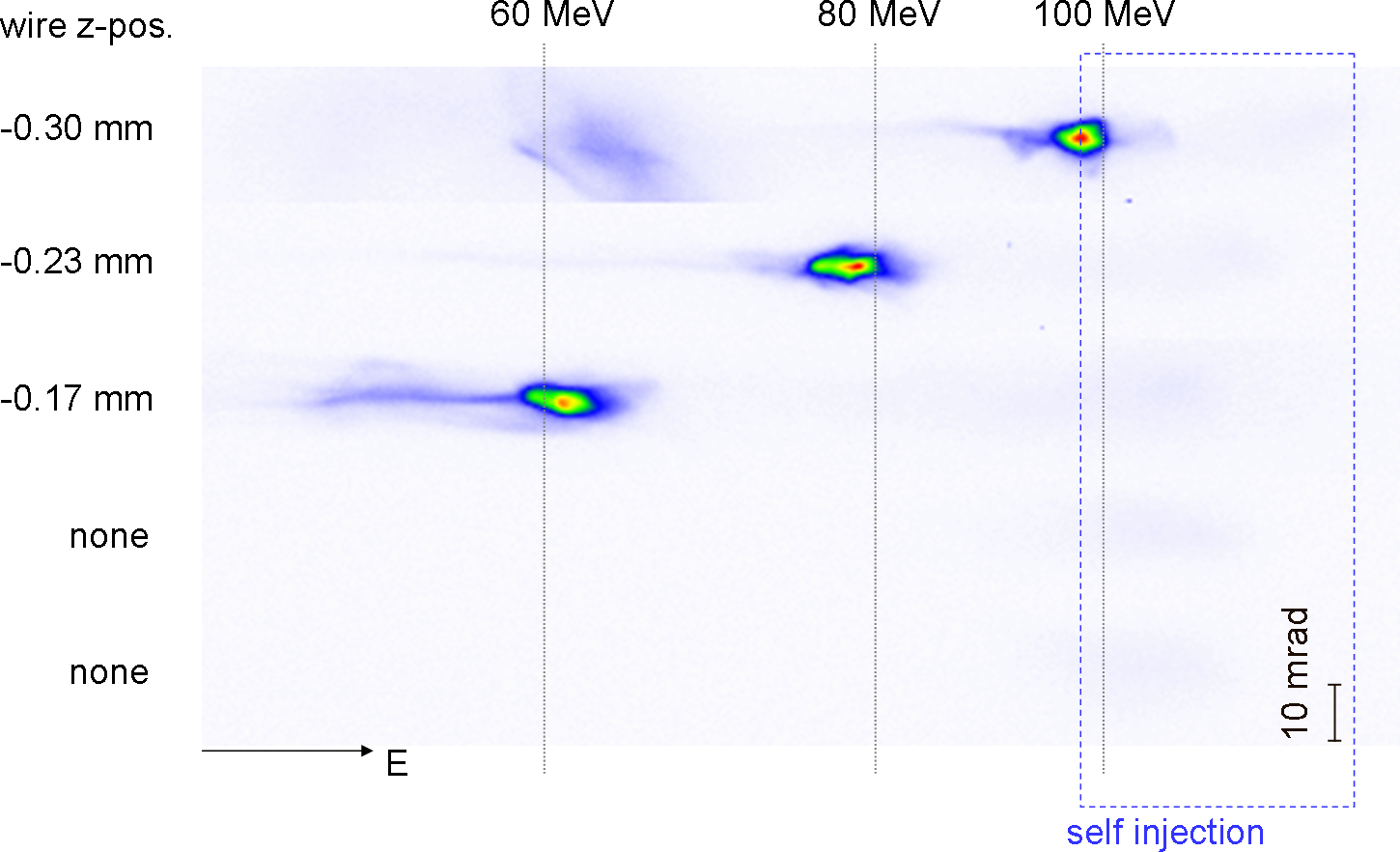}
	\caption{Example spectra  with comparable charge and variable wire position as recorded on the Lanex screen using $9.5\,$bar backing pressure. Besides the rather strong peak when the wire is present, a weak background self-injection can be seen in all spectra.}
	\label{fig:spectracompilation}
\end{figure}

\begin{figure}
	\centering
		\includegraphics[width=\columnwidth]{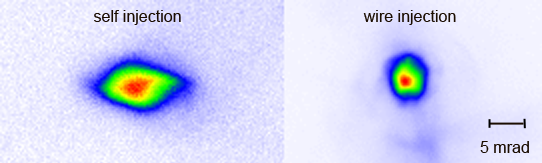}
	\caption{Beam profile measurements acquired with helium illustrate different eccentricities for the two cases without wire on the left and with wire on the right, shown on an equal lateral and normalized color scale.
	\label{fig:He_beamprofiles}}
\end{figure}

Example spectra can be seen in Fig. \ref{fig:spectracompilation}, showing the spectral range from $43\,$MeV to infinity. A relative spectral spread $\frac{\Delta E}{E} \leq 4\,$\% can be calculated. Note however, that spectrometer dispersion and divergence have not been deconvoluted here. In fact, $4\,$mrad FWHM (see Fig. \ref{fig:spectracompilation}) produces an apparent $\frac{\Delta E}{E} \approx 4\,$\% (FWHM) at $100\,$MeV, thus the real relative spectral spread is below what can be resolved with this particular spectrometer but less than a few percent. Electron beam mean energies are generally lower with the wire. The effect of the wire is threefold: It injects a charge $\sim 10 \,$ times higher than that available without wire while at the same time providing beams with clean quasi-monoenergetic spectra and reduced divergence, thus brightness is increased dramatically. A weak self-injected background charge can be identified in most of the shots. Within limits, energy tuning becomes possible by altering the $z$-position of the wire as illustrated in Fig. \ref{fig:spectracompilation}. From this a field gradient $\sim 250\,$GV/m and an acceleration length of $\approx 0.4\,$mm may be estimated, indicating that acceleration for the wire-injected beams affectively only takes place in stage III. The background charge, which is higher in energy, must therefore result from injection in an earlier stage, or resemble an injected dark current exposed to higher acceleration fields. If we assume this background not to be injected much earlier than the wire-injected charge, this indicates that acceleration of the wire-injected charge takes place in a later plasma wave period. This is supported by beam profile measurements that show an ellipticity in the beam divergence for the self-injected beams only, as illustrated in Fig. \ref{fig:He_beamprofiles}, which can be understood as an effect due to the interaction of the injected electrons with the laser field inside the first plasma wave period \cite{Mangles2006}.

\begin{figure}
	\centering
		\includegraphics[width=\columnwidth]{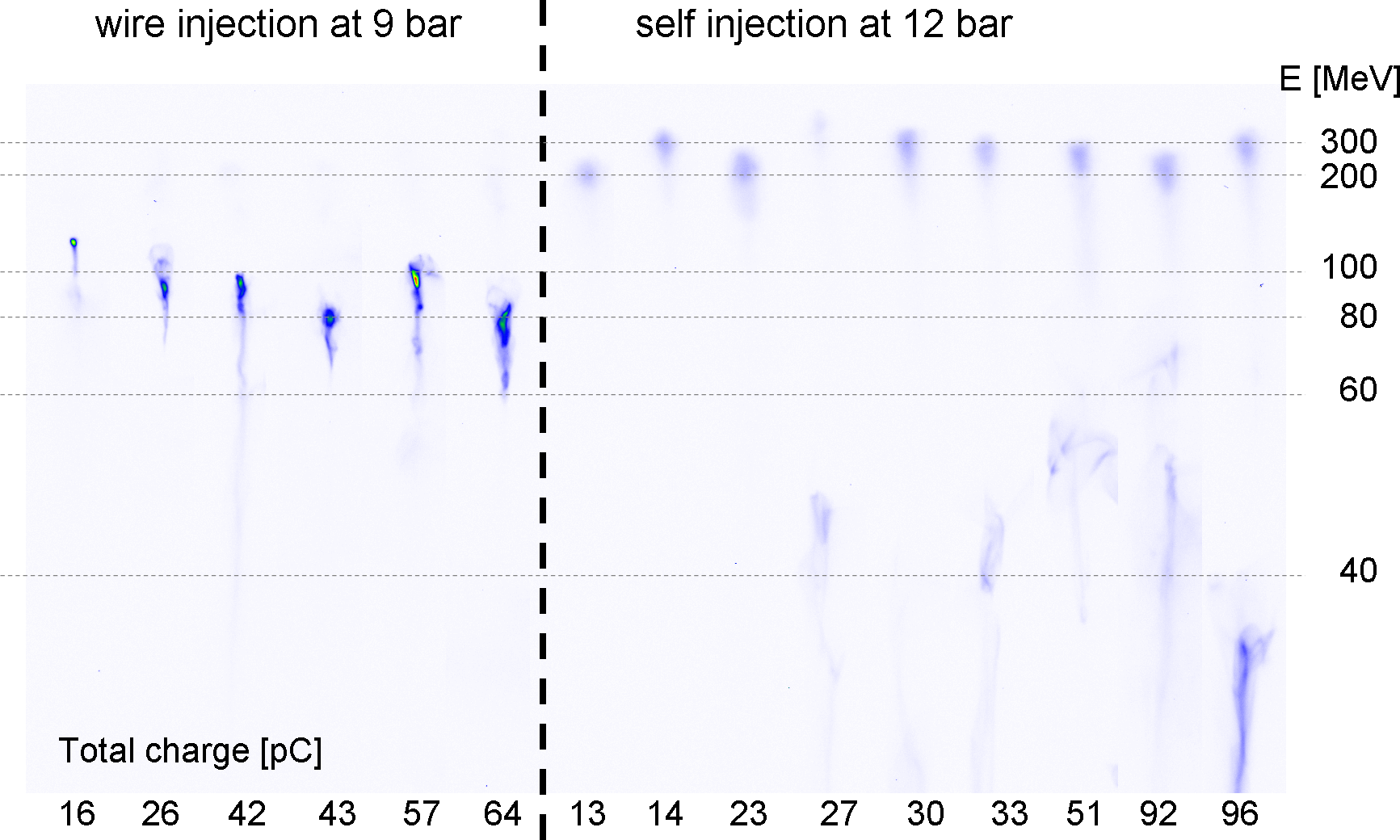}
	\caption{Example spectra of beams with variable charge and fixed wire position; left: wire injection at $9\,$bar backing pressure; right: self-injection at $12\,$bar backing pressure to compensate for the charge increase in the wire injection case as indicated in Fig. \ref{fig:beamfeatures}. Each group has been sorted according to integrated charge}
	\label{fig:spectracompilationsc}
\end{figure}

Figure \ref{fig:spectracompilationsc} shows spectra with and without wire at backing pressures that result in comparable beam charges for both cases. Wire injection at $9\,$bar is thus compared to self-injection at $12\,$bar. Note that low-energy artifacts, carrying a significant amount of charge, appear $> 20\,$pC in the spectra of self-injected beams, while spectra of wire-injected beams are cleaner. The tendency of decreasing peak energy with increasing charge due to beam loading \cite{Rechatin2009} is clearly visible in the wire injection case and indicates that injection probably occurs at the same $z$-position.\\

The 3D fully relativistic parallel PIC code \textit{ELMIS} \cite{ELMIS} was used to investigate the physical mechanisms in the modulated density during laser propagation. In the simulation $140\,$attoseconds corresponded to one time step and an $80 \times 80 \times 80 \, \mu$m$^3$ box was represented by $1024\times 256\times 256$ cells. The ions (H$^{+}$) were mobile. During the simulation the average number of virtual particles was 1 billion. Laser parameters were taken from the experiment. As the resolution for the measured plasma densities was limited to $\sim 100\,\mu$m, the exact distribution is unknown. Still, the interferometric data does provide useful information about densities and lengths of each section while steeper gradients such as shock fronts remain concealed. In combination with theoretical considerations and fluid simulations by Wang \textit{et al.} \cite{Wang2012}, a profile resembled by the broken line function in Fig. \ref{sima} is very likely and thus used for the simulations.  This density distribution is fully consistent with the acquired interferometric data. It should be noted though, that estimated gas jet temperatures in the experiment are in the range $10 - 50\,$K only, and are thus one order of magnitude lower than those simulated by Wang \textit{et al.} \cite{Wang2012}. Moreover, in our case, the distance between wire and plasma channel was about three times larger than what was presented in that particular paper. Additionally, those simulations were carried out with helium while we used hydrogen, which at these low temperatures requires a different equation of state.

With the proposed density distribution, simulations show that when traversing region I, the laser pulse gets focused transversely and generates a highly nonlinear plasma wave, which does not reach breaking and thus facilitates neither longitudinal nor transverse self-injection of electrons. In line with previous studies \cite{Bulanov1998, suk.prl.2001, hemker.prst.2002} at the density downramp, which is the expansion fan originating from the wire, the cavities of the nonlinear plasma wave rapidly expand behind the laser pulse and thereby catch electrons accumulated between the buckets. In region II these electrons form an electron bunch. At the entry to region III, the cavity size shrinks again. With the assumed shallow inward gradients however, injected electrons are dephased after the transition and are not accelerated further. The following outward density shock wave enables a second injection of a new bunch into the second plasma wave period, which is accelerated throughout region III. 

The initial dephasing and renewed injection of electrons upon entering stage III, followed by a rapid acceleration over no more than half a millimeter until the end of the gas jet, explains the rather short acceleration distances that can be derived from the field estimates related to Fig. \ref{fig:spectracompilation}. This also leads to the observed lower final energy of the electrons compared to the self-injection case. The localized injection results in highly monoenergetic beams both in simulation and experiment. When increasing the height above the wire, a decrease in gradient strengths together with an extension of region II explains the reduced probability for the production of electron beams when operating at too large a distance. That the electron are accelerated in the second plasma wave period, isolated from the laser pulse, as suggested by simulations and indicated experimentally, may as well explain the reduced divergence.

In conclusion, a wire injection scheme has successfully been demonstrated as an alternative to some more complex set-ups facilitating controlled injection. Beam features include a reduction in divergence by $25\,$\% and an increase in bunch charge by up to one order of magnitude if compared to beams of electrons accelerated in the quasi-monoenergetic, nonlinear, self-injection regime. Their spectra are tunable and show less than a few percent relative spread. Simulations confirm the experimental findings with respect to external injection control, acceleration in the second plasma wave period and resulting spectral distribution.\\

We acknowledge the support of the Swedish Research Council, the Knut and Alice Wallenberg Foundation, the Lund University X-Ray Centre, the European Research Council contract number 204059-QPQV, the Swedish Research Council contract number  2010-3727 and the Swedish National Infrastructure for Computing.

\end{document}